\documentclass[a4paper,aps,pra,preprintnumbers]{revtex4-2}
\usepackage[utf8]{inputenc}
\usepackage[english]{babel}
\usepackage[T1]{fontenc}
\usepackage{amssymb,amsfonts,amsmath,mathtext,enumerate,float,dsfont}
\usepackage{graphics,graphicx,epsfig,epstopdf}
\usepackage{caption}
\usepackage{cmap}
\usepackage{multirow}
\usepackage{indentfirst}
\usepackage[usenames]{color}
\usepackage{amsthm}
\usepackage{xcolor}

\begin{document}

\title{Particle-loss error correction using $2^l$-step states generated via particle-number-resolving measurement schemes}

\author{S. B. Korolev}
\author{E. A. Nesterova}
\affiliation{St.Petersburg State University, Universitetskaya nab. 7/9, St.Petersburg, 199034, Russia}

\begin{abstract}
We present an iterative scheme for generating finite superpositions of Fock states whose photon numbers differ by $2^l$ ($2^l$-step states). We derive an upper bound on the probability of generating such states. Based on the proposed class of states, we construct several bosonic encodings and compare their performance in correcting particle-loss errors. Our results identify the most efficient encoding, which achieves a higher channel fidelity than previously studied bosonic encodings.
\end{abstract}
\maketitle
\section{Introduction}
Photons are fundamental carriers of quantum information. A major obstacle to quantum communication, however, is that photons are rapidly absorbed in a communication channel such as an optical fiber. For this reason, quantum error correction is an essential component of quantum information processing. Protection against unavoidable noise is necessary during the transmission, storage, and processing of quantum information.

Quantum error correction protects quantum information through redundancy: information is encoded so that the original state can be reconstructed even if part of it is lost \cite{Nielsen_Chuang_2010}. Bosonic codes occupy a special place among such methods. In contrast to the traditional approach, where a logical qubit is encoded in multiple physical qubits \cite{Peres1985,Shor1995,Steane1996}, bosonic codes exploit the infinite-dimensional Hilbert space of a single quantum oscillator \cite{Leghtas2013,Albert2018,Terhal_2020,Wallraff2004,Terhal2016,Wasilewski2007,Glancy2004}. This avoids the need for complex systems of interacting qubits and offers practical implementation advantages.

Several classes of bosonic codes have been proposed. Protection against particle-loss errors can be achieved using Gottesman–Kitaev–Preskill (GKP) codes \cite{Gottesman2001,Vasconcelos2010}, codes based on Schrödinger cat states \cite{Cochrane1999,Bergmann2016,Mirrahimi_2014}, and binomial codes \cite{Schlegel2022,Korolev2024_eror_cor, Binomial_state}. Each of these classes offers its own advantages, but they share a common problem: ideal code states cannot be prepared exactly in an experiment. Only approximate versions of these states are experimentally accessible \cite{Konno2024,Kudra2022,Sychev2017,Ourjoumtsev_cat2007,Polzik2006, Huang2015, Ulanov2016, Gerrits2010, Takahashi2008,Thekkadath2020}, which can significantly reduces the efficiency of error correction. This motivates the search for bosonic codes that combine effective error protection with realistic experimental preparation.

Ref. \cite{2-step} proposed a bosonic code whose codewords are generated by measuring the particle number in a two-mode Gaussian state and can therefore be prepared experimentally. The logical codewords are finite superpositions of Fock states whose numbers differ by multiples of two (2-step states). Such states provide protection against single particle-loss errors, but are not suitable for correcting two or more particle-loss errors.

In this work, we develop this approach and consider an iterative scheme in which the step of the superposition of Fock states doubles at each iteration. We analyze the probability of state generation at each iteration and investigate how increasing the step affects the efficiency of particle-loss error correction. This analysis identifies the states in this family that are optimal in terms of channel fidelity and experimental generation probability. Using $4$-step states, we search for the optimal encoding and compare it with binomial encoding, encoding using four‑component Schrödinger cat states, and encoding based on Fock states with a step of two.

\section{Quantum channel with particle-loss errors and error correction codes}

Before describing the states used to protect quantum information against particle-loss errors during transmission through a quantum channel, we first define the action of such a channel on quantum states. In such a channel, the evolution of a quantum state described by the density matrix $\hat{\rho}$ is governed by the Lindblad master equation \cite{Lindblad1976}:
\begin{align} \label{eq_lin}
\partial_{t} \hat{\rho}(t)=\kappa_{1} \mathcal{D}[\hat{a}] \hat{\rho}(t), 
\end{align}
where $\kappa_{1}$ is the particle-loss rate, and the superoperator is defined as
\begin{align}
\mathcal{D}[\hat{a}] \hat{\rho}(t) & =\hat{a} \hat{\rho}(t) \hat{a}^{\dagger}-\frac{\hat{a}^{\dagger} \hat{a} \hat{\rho}(t)+\hat{\rho}(t) \hat{a}^{\dagger} \hat{a}}{2}.
\end{align}
The annihilation $\hat{a}$ and creation operators $\hat{a}^{\dagger}$ obey the canonical commutation relation  $\left[\hat{a},\hat{a}^{\dagger}\right]=1$.  The solution to equation (\ref{eq_lin}) can be expressed in the Kraus decomposition form \cite{KRAUS1971311}:
\begin{align} \label{evol_los}
\hat{\rho}(t)\equiv \mathcal{N}\left[\gamma\right]\left(\rho(0)\right)=\sum_{j=0}^{\infty} \hat{K}_{j} \hat{\rho}(0) \hat{K}_{j}^{\dagger},
\end{align}
where the Kraus operators  $\lbrace\hat{K}_i\rbrace$  for the loss channel are given by \cite{MUeda_1989,Lee_1994,Chuang_1997}:
\begin{align} \label{Krauss_loos}
    \hat{K}_k=\sqrt{\frac{\gamma^k}{k!}}(1-\gamma)^{\hat{n}/2}\hat{a}^k.
\end{align}
Here $\hat{n}=\hat{a}^{\dagger}\hat{a}$ is the particle number operator, and $\gamma=1-e^{-\kappa_1 t}$  is the dimensionless damping parameter, which ranges from $0$ to $1$.

Quantum information must be encoded to protect it from losses in this channel. A bosonic quantum code encodes the two logical qubit states $|\overline{0}\rangle$ and $|\overline{1}\rangle$ into a quantum oscillator. These states are referred to as codewords (CWs). The code space  $\mathcal{H}_L=\mathrm{span}\lbrace |\overline{0}\rangle,|\overline{1}\rangle \rbrace$ is a subspace of the oscillator's infinite-dimensional Hilbert space.

The CWs $|\overline{0}\rangle$ and $|\overline{1}\rangle$ can be used for error correction in channels described by a set of Kraus operators $\lbrace\hat{K}_i\rbrace$ if the following condition is satisfied \cite{Knill_1997}: 
\begin{align} \label{KL_ideal}
    \left|\left|\hat{P}\hat{K}_i^\dagger \hat{K}_j \hat{P}-\lambda_{ij}\hat{P}\right|\right|<\varepsilon,
\end{align}
where $\varepsilon>0$ is an arbitrarily small constant, $\hat{P}=|\overline{0}\rangle\langle \overline{0}|+|\overline{1}\rangle \langle \overline{1}|$ is the projector onto the codeword subspace and $\lambda_{ij}=\lambda_{ji}^*$ are the elements of a Hermitian
matrix independent of the CWs. The error-correction performance improves as $\varepsilon$ decreases.  In the limit $\varepsilon =0$, the code can correct all errors in the channel. In this case, the code is referred to as perfect. However, no perfect code for realistic quantum channels has been found to date.

To quantify the performance of different CWs in quantum error correction, one typically evaluates the channel fidelity. The channel fidelity characterizes how close the state recovered after transmission through the channel is to the original state: $ F=F\left(\mathcal{R}\circ \mathcal{N} \left(|\psi\rangle \langle \psi|\right),|\psi \rangle  \right)$.  Here, $\mathcal{R}$ denotes the recovery channel, which recovers the quantum information after transmission through the noisy channel $\mathcal{N}$.  We adopt the transpose channel \cite{Barnum2002,Ng2010} , one of the channels close to optimal. This channel has been used in many studies \cite{Chuang1997,Kosut2009,TC} due to the relative simplicity of the calculations and the absence of fundamental limitations on its practical implementation \cite{Biswas2024}.

The fidelity of a quantum channel in which the error process is followed by recovery via the transpose channel is given by \cite{Zheng2024}:
\begin{align} \label{fid_chan}
\mathcal{F}=\frac{1}{4}\left\|\operatorname{Tr}_{L} \sqrt{M}\right\|_{F}^{2}, 
\end{align}
where $\left|\left|\cdot\right|\right|_F$ denotes the Frobenius norm, $\left(\operatorname{Tr}_{L} B\right)_{l, k}=\sum_{\mu} B_{[\mu l],[\mu k]}$ represents the partial trace over the code space indices, and $M$ is the quantum error correction (QEC) matrix. This matrix is defined as
\begin{align} \label{QEC_matrix}
    M_{[\mu l],[\nu k]}:=\left\langle\overline{\mu}\right| \hat{K}_{l}^{\dagger} \hat{K}_{k}\left|\overline{\nu}\right\rangle,
\end{align}
where $|\overline{\mu}\rangle$, $|\overline{\nu}\rangle$ are orthogonal CWs, and $\left\{\hat{K}_{i}\right\}$ are the Kraus operators of the error channel $\mathcal{N}$.

In this work, we use the channel fidelity defined in Eq. (\ref{fid_chan}) to evaluate different states generated within a specific scheme and assess their suitability as codes for correcting particle-loss errors. We now proceed to describe the set of states under investigation.

\section{Set of generated states }
\subsection{General generation scheme }
\begin{figure}[H]
    \centering
    \includegraphics[width=0.4\linewidth]{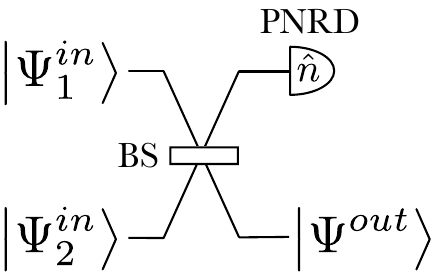}
    \caption{Scheme for generating states used for error-correction codes. In the figure: $|\Psi_1^{in}\rangle$ and $|\Psi_2^{in}\rangle$ are two input quantum states, BS is a beam splitter, PNRD is a particle-number-resolving detector, and $|\Psi^{out}\rangle$ is the output state.}
    \label{fig:scheme}
\end{figure}

To generate states suitable for protecting quantum information from particle-loss errors, we use the scheme shown in Fig.\ref{fig:scheme}.
\noindent In this scheme, two quantum states $|\Psi_1^{in}\rangle$ and $|\Psi_2^{in}\rangle$ are mixed at a beam splitter with amplitude transmission coefficient $t$. This produces a two-mode state, one mode of which is measured by a particle-number-resolving detector (PNRD). For definiteness, we assume that the detector has measured $n$ particles. The second mode is then in an output state, which we use as CWs for error correction.

\subsection{Generation of 2-step states}
If squeezed vacuum states are used as input states, $|\Psi_1^{in}\rangle=\hat{\mathcal{S}}(\rho_1)|0\rangle$ and $|\Psi_2^{in}\rangle=\hat{\mathcal{S}}(\rho_2)|0\rangle$, where $\hat{\mathcal{S}}\left(\rho_i\right)=\exp\left[\frac{\rho_i}{2}\left(\left(\hat{a}^\dag\right)^2-\hat{a}^2\right)\right]$ is the squeezing operator, and $\rho_1,\rho_2 \in \mathbb{R}$ are the squeezing parameters. The output state then has the form  \cite{Korolev_2024}:
\begin{align} \label{out_rw}
    \left|\Phi_n^{1} \left(r,z\right)\right \rangle = \hat{\mathcal{S}}\left(r\right)\sum _{k=0}^{\left\lfloor \frac{n}{2}\right\rfloor } \frac{ (-z)^k}{2^{k}\,k!} \sqrt{\frac{n!}{(n-2 k)! \, _2F_1\left(\frac{1-n}{2},-\frac{n}{2};1;z^2\right)}} \left| n-2 k\right \rangle,\end{align}
where $n$ is the number of detected particles in the scheme, $|k\rangle$ denotes a Fock states, and $_2F_1\left(\frac{1-n}{2},-\frac{n}{2};1;\left| z\right| ^2\right)$ is a hypergeometric function. The parameters of the generated state are related to those of the generation scheme as follows  \cite{Korolev_2024,Korolev2024_fock,Bashmakova_2023}:
   \begin{align}
       &z=-\frac{ (1-t) \sinh (2 \rho_1)+t \sinh (2
   \rho_2)}{2 \sinh^2(\rho_1-\rho_2) (1-t) t},\\
  & \cosh (r)=\frac{1}{\sqrt{(1-t \tanh (\rho_1)-(1-t) \tanh (\rho_2)) (1+t \tanh
   (\rho_1)+(1-t) \tanh (\rho_2))}}.
   \end{align}
The resulting state is a finite superposition of squeezed Fock states and depends on three parameters: $r$, $z$, and $n$. The parameter $r$ determines the squeezing of the Fock states, while $z$ controls the superposition coefficients.

It was shown in Ref. \cite{2-step} that the states in Eq. (\ref{out_rw}) can be expanded in the Fock basis and contain either only even or only odd Fock states. This means that the numbers of neighboring Fock states in the expansion (\ref{out_rw}) differ by two ($|m+2-m|=2$). We refer to finite superpositions of Fock states whose numbers differ by multiples of $N$ as $N$-step states.

The definite parity of these states makes them suitable for correcting particle-loss errors. A single particle-loss error changes the parity of a state (\ref{out_rw}), allowing the error to be detected and corrected. The loss of two particles does not change the parity, making this error indistinguishable from the absence of an error. Therefore, states (\ref{out_rw}) are suitable only in quantum channels in which the probability of losing two or more particles is low.

\subsection{Generation 4-step states}
Next, we search for states that can correct more particle-loss errors. We send the states (\ref{out_rw}) with the numbers $n_1$ and $n_2$ and the parameters $z_1$, $r_1$ and $z_2$, $r_2$ to the input of the scheme shown in Fig.  \ref{fig:scheme}. We therefore consider the case when $|\Psi_1^{in}\rangle=\left|\Phi_{n_1}^{1} \left(r_1,z_1\right)\right \rangle$ and $|\Psi_2^{in}\rangle=\left|\Phi_{n_2}^{1} \left(r_2,z_2\right)\right \rangle$. 
This scheme reduces to a scheme in which two squeezed vacuum states are used to generate two states (\ref{out_rw}), which are then sent to the scheme shown in Fig. \ref{fig:scheme}.

We mix the states $\left|\Phi_{n_1}^{1} \left(r_1,z_1\right)\right \rangle$ and $\left|\Phi_{n_2}^{1} \left(r_2,z_2\right)\right \rangle$ at a beam splitter with an amplitude transmission coefficient $t$. Measuring $m$ particles at the PNRD, we obtain a state in the unmeasured mode of the form:
\begin{multline} \label{out_4step}
   |\Phi_{n_1,n_2,m}^2(r_1,r_2,z_1,z_2) \rangle \\ 
   \propto \left \langle m \right |\hat{BS}(t) \left|\Phi_{n_1}^1 (r_1,z_1)\right\rangle\left|\Phi_{n_2}^1(r_2,z_2)\right\rangle \\ 
   \propto \hat{\mathcal{S}} \left(R\right)\left( \sum_{k=0}^{\lfloor \frac{n_1+n_2+m}{2} \rfloor} \alpha_{n_1+n_2+m-2k}   |n_1+n_2+m-2k\rangle \right).
\end{multline}
The superposition coefficients are given by
\begin{align}
 \alpha_k= \sqrt{\frac{2^k}{k!\,c^{k+1}}}  \left.
\frac{\partial^{\,m+n_1+n_2} }
     {\partial u_1^{m}\,\partial u_2^{n_1}\,\partial u_3^{n_2}}\left(Q_k(u_1,u_2,u_3)  \exp \left(P(u_1,u_2,u_3)\right)\right)
\right|_{u_1=0,\;u_2=0,\;u_3=0},
\end{align}
where the polynomials take the form
\begin{align}
  Q_k(u_1,u_2,u_3)=  \left(\sqrt{1-t^2}  \cosh \left(r_2\right) u_2+t  \cosh \left(r_1\right)u_3-t \sqrt{1-t^2}  \sinh \left(r_1-r_2\right) u_1\right)^k,
\end{align}
\begin{multline}
    P(u_1,u_2,u_3)=u_1^2 \sinh \left(R-r_1-r_2\right)-u_2^2 \left(c z_1-t^2 \sinh \left(R+r_1-r_2\right)\right)\\
    -u_3^2 \left(c z_2-\left(1-t^2\right) \sinh \left(R-r_1+r_2\right)\right)\\
    -2 u_1
   \left(\sqrt{1-t^2} u_3 \cosh \left(R-r_1\right)-t u_2 \cosh \left(R-r_2\right)\right)-2 u_2 u_3\, t \sqrt{1-t^2}  \sinh (R),
\end{multline}
We also introduce the following abbreviation: 
\begin{align}
    c=\sqrt{\cosh ^2\left(r_2\right)+2 t^2 \sinh \left(r_1\right) \sinh \left(r_1-r_2\right) \cosh \left(r_2\right)-t^4\sinh ^2\left(r_1-r_2\right)}.
\end{align}

The squeezing parameter of the generated state is related to the parameters of the input states and the beam splitter transmission coefficient as follows:
\begin{align}
    e^{2 R}=\frac{e^{r_1} \cosh \left(r_2\right)-t^2 \sinh \left(r_1-r_2\right)}{e^{-r_1} \cosh \left(r_2\right)+t^2 \sinh \left(r_1-r_2\right)}.
\end{align}

As follows from Eq. (\ref{out_4step}), for arbitrary values of the parameters $r_1$, $z_1$, $r_2$, $z_2$, $t$, $n_1$, and $n_2$, the generated state is a finite superposition of squeezed Fock states with definite parity. Thus, in the general case, this scheme also generates 2-step states.

We now consider a special case of this scheme that allows us to generate states with a larger step. To this end, we set $t=1/\sqrt{2}$, $r_2=-r_1=\tilde{r}$, $z_1=-z_2=\tilde{z}$ and $n_2=n_1=\tilde{n}$. This case corresponds to the mixing, at a symmetric beam splitter, of two states (\ref{out_rw}) that are squeezed in orthogonal directions and have the same indices $n$. As a result, the squeezing parameter of the generated state vanishes: $R=0$. The resulting state then takes the form:
\begin{align} \label{out_4step_1}
 |\Phi_{\tilde{n},m}^2(\tilde{r},\tilde{z})\rangle=  \sum _{k=\max \left(\lceil \frac{2 \tilde{n} - m}{4} \rceil,0\right)}^{\tilde{n}}  \frac{1}{\sqrt{\mathcal{N}_2}}\tilde{\alpha}_{k}(\tilde{r},\tilde{z})|m-2\tilde{n}+4k\rangle,
\end{align}
where the superposition coefficients are defined as follows:
\begin{align} \label{coef_4k}
\tilde{\alpha}_{k}(\tilde{r},\tilde{z})=\frac{(-1)^k (2 k)! \left(2 (\tilde{n}-k)\right)!
   \left(\frac{\text{sech}^2\left(\tilde{r}\right)}{\tanh \left(\tilde{r}\right)+\tilde{z}}\right)^{2 k}
   \, _2\tilde{F}_1\left(-m,1+2 k;2 \tilde{n}-m-2 k+1;\frac{\tilde{z}+\tanh \left(\tilde{r}\right)}{\tilde{z}+\coth
   \left(\tilde{r}\right)}\right)}{k! \left(\tilde{n}-k\right)! \sqrt{\left(m-2 \tilde{n}+4
   k\right)!}},
\end{align}
and $\mathcal{N}_2=\sum _{k=\max \left(\lceil \frac{2 \tilde{n} - m}{4} \rceil,0\right)}^{\tilde{n}} \tilde{\alpha}_k^2$ is the normalization factor, and $_2\tilde{F}_1\left(a,b;c ;x\right)$ denotes the regularized hypergeometric function.

It follows from Eq. (\ref{out_4step_1}) that the generated states are 4-step states. This means that, in this particular case, the scheme generates states with a step twice as large as that in the scheme with two squeezed vacuum states. The generated state also depends on three parameters of the input states: $\tilde{r}$, $\tilde{z}$ and $\tilde{n}$, as well as on the number of detected particles $m$. This number affects the parity of the generated state, as well as which Fock state numbers are present in the superposition. The maximum Fock state number participating in the superposition is $m+2n_1$, and the minimum number is either zero or $\lceil \frac{2 \tilde{n} - m}{4} \rceil$. For example, for $\tilde{n}=4$ and $m=1$, we obtain a state of the form $\tilde{c}_2 |1\rangle+\tilde{c}_3 |5\rangle +\tilde{c}_4|9\rangle$, and for $\tilde{n}=2$ and $m=4$, we obtain the state $\tilde{c}_0| 0\rangle+\tilde{c}_1 |4\rangle+\tilde{c}_2 |8\rangle$.

\subsection{Generation $2^{l}- $step states}
Now we generalize the approach we have considered to generate  $2^{l+1}$-step states. To generate such states, we also use the scheme shown in \ref{fig:scheme}, with a symmetric beam splitter. As the first input state, we take a $2^l$-step state:
\begin{align}
    |\Psi_{1}^{in} \rangle= \sum _{k_1=0}^{N_2} \beta_{j + 2^{l}\,k_1}|j+2^{l} k_1\rangle,
\end{align}
where $N_2$ and $j$ are non-negative integers that define the maximum and minimum Fock states involved in the expansion, and $2^l$ determines the step between Fock states. As the second state, we consider a state that differs from the first only by a rotation of$\pi/(2^l)$:
\begin{align}
    |\Psi_{2}^{in} \rangle= e^{\frac{i\pi\hat{n}}{2^l}}|\Psi_{1}^{in} \rangle=e^{\frac{i\pi j}{2^l}} \sum _{k_1=0}^{N_2} e^{i \pi k_1}\beta_{j + 2^l\,k_1}|j+2^l\, k_1\rangle. 
\end{align}
Such a transformation can be implemented in optics by inserting a glass plate of appropriate thickness into the second channel.
After mixing at a symmetric beam splitter and measuring $m$
 particles at the detector, we obtain a state of the form:
\begin{align}
  |\Phi ^l_{out}\rangle= \frac{\sqrt{m!}}{2^{j+2^l N_2}\sqrt{\mathcal{N}}_{2^l}} \begin{cases}
           \sum \limits_{k=0}^{N_2}  \tilde \beta_{2k} |2j +2^{l+1} N_2-m-2^{l+1}\,k\rangle, \quad \text{if}\quad m\, \text{-- even}\\
        \sum \limits_{k=0}^{N_2}  \tilde \beta_{2k-1}|2j +2^{l+1} N_2-m-2^{l}(2\,k-1)\rangle, \quad \text{otherwise}
    \end{cases},
\end{align}
where $\mathcal{N}_{2^l}$ is the normalization factor, and the superposition coefficients are determined by this sum:
\begin{multline} \label{coer_sup_gen}
    \tilde \beta_k=\sum_{\tilde{k}=\max \left(0,N_2-k\right)}^{\min \left(N_2,2 N_2-k\right)}(-1)^{\tilde{k}} \beta_{j+2^{l}\tilde{k}} \beta_{j+2^{l+1}N_2-2^l\tilde{k}-2^l k}\sqrt{ \frac{2^{2^{l}k}{\left(j+2^{l+1}N_2-2^l \tilde{k}-2^lk\right)!}  \, }{{\left(j+2^l\tilde{k} \right)!} {\left(2 j+2^{l+1}N_2-m-2^lk\right)!}}} \\
   _2\tilde{F}_1\left(-j-2^l
   \tilde{k},-2 j+m-2^l \left(2 N_2-k\right);-j+m-2^l \tilde{k}+1;-1\right).
\end{multline}
As follows from Eq.\ref{coer_sup_gen}, the scheme generates $2^{l+1}$-step states. Therefore, the scheme generates states with twice step of the input states. In particular, by sending two states (\ref{out_4step_1}) with a step of $2^l=4$ to the input of the scheme, we obtain states with a step of $2^{l+1}=8$.

\begin{figure}[H]
    \centering
    \includegraphics[width=0.5\linewidth]{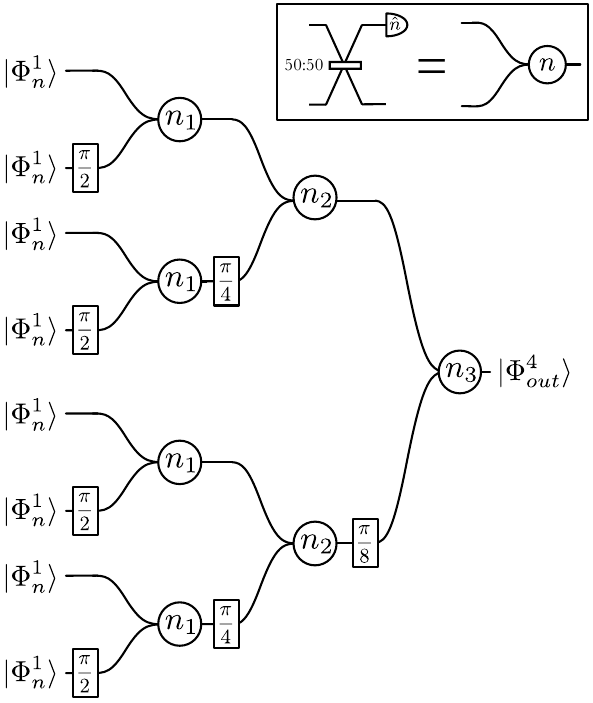}
    \caption{Iterative scheme for generating $16$-step states. In the figure, rectangles labeled $\pi/k$ represent rotation transformations, while circles labeled $n_i$ denote a scheme with a symmetric beam splitter and measurement of the photon number $n_i$. The state $|\Phi_n^1\rangle$ is a $2$-step state, and $|\Phi_{out}^4\rangle$ is the output $16$-step state.}
    \label{fig:scheme_iteration}
\end{figure}

Therefore, the states (\ref{out_rw}) and the symmetric-beam-splitter scheme shown in Fig.  \ref{fig:scheme} allow us to generate  $2^{l+1}$-step states through successive iterations. For example, Fig.  \ref{fig:scheme_iteration} shows an iterative scheme for generating $16$-step states $|\Phi_{out}^4 \rangle$.
\noindent In this scheme, eight identical states  $|\Phi _n^1\rangle$  are transformed iteratively by schemes with symmetric beam splitters and particle-number measurements, yielding a 16-step state. Using such iterative schemes, we can, in theory, generate states with various steps of $2^l$. As $l$ increases, such states should correct particle-loss errors more effectively. We will discuss this in more detail later.

As can also be seen from Fig.  \ref{fig:scheme_iteration}, the generated states depend on two parameters, $\tilde{r}$ and $\tilde{z}$. Indeed, the superposition coefficients in Eq. (\ref{coer_sup_gen}) are expressed in terms of the coefficients $\beta_k$ of the input states, while the resource states $|\Phi^1_n(\tilde{r},\tilde{z})\rangle$ depend on $\tilde{r}$ and $\tilde{z}$. Consequently, any generated $2^l$-step state depends on these two parameters, and varying them allows one to control the superposition coefficients of the generated states.

\subsection{Probability of generating states at different iterations of the scheme }
The generation of the proposed states is probabilistic, as it results from measuring a certain number of particles $m$. This probability decreases with each iteration of the scheme. The probability of generating states (\ref{out_rw}), which serve as the basic elements for $2^l$-step states, is given by the following expression:
\begin{align} \label{prob_1_step}
    P_1(n;a,z)=\frac{2(a-1)^n}{(a+1)^{n+1}| 1-z| ^{n+\frac{1}{2}}} \sqrt{a(| 1-z| -1)+1} \,
   _2F_1\left(\frac{1-n}{2},-\frac{n}{2};1;| z| ^2\right),
\end{align}
where $a>0$ is a free parameter. The parameter $a$ can be adjusted to maximize the probability of generating states with a specific parameter $z$. Details on maximizing the probability for various states can be found in Refs.\cite{Korolev_2024}. Here, we focus only on the maximum achievable probabilities. This expression for the probability is bounded above \cite{Fiurasek_2026_1} by the value $\frac{n^{n}}{(n+1)^{n+1}}$. This quantity shows that as the number of detected particles $n$ increases, the probabilities of state generation decrease. Thus, the probability of generating a state when measuring one particle does not exceed $25\%$, and when measuring $m=20$ particles it is no more than  $2\%$.

The probability of generating the state $|\Phi _{\tilde{n},m}^2\rangle$ is the product of the probabilities of generating two input states (\ref{prob_1_step}) and the probability of measuring $m$ particles in the scheme shown in Fig.  \ref{fig:scheme},whose input states are $|\Phi _{\tilde{n}}^1\rangle$. Consequently, the probability of generating $4$-step states is given by the following expression:
\begin{multline} \label{prob_2_step}
  P_2(\tilde{n},m;\tilde{r},\tilde{z},a_1,a_2)= P_1(n_1;a_1,\tilde{z})P_1(\tilde{n};a_2,-\tilde{z})\\
  \times\frac{ 1  }{\cosh^2(\tilde{r}) m! \, _2F_1\left(\frac{1-\tilde{n}}{2},-\frac{\tilde{n}}{2};1;\tilde{z}^2\right)^2}  \left(\frac{\coth (\tilde{r})+\tilde{z}}{\tilde{z} \coth (\tilde{r})+1}\right)^{2 m} \left(\frac{\tilde{z} \cosh
   (\tilde{r})+\sinh (\tilde{r})}{\sqrt{2}}\right)^{4\tilde{n}}\\
  \times \sum _{r=\max \left(\lceil \frac{2 \tilde{n} - m}{4} \rceil,0\right)}^{\tilde{n}} \left(\frac{(2 r)! (2\tilde{n}-2 r)!}{r! (\tilde{n}-r)!}\right)^2 \left(\frac{\text{sech}^2(\tilde{r})}{\tanh (\tilde{r})+\tilde{z}}\right)^{4 r}\frac{  \, _2\tilde{F}_1\left(-m,2 r+1;-m+2
  \tilde{n}-2 r+1;\frac{\tilde{z}+\tanh (\tilde{r})}{\tilde{z}+\coth (\tilde{r})}\right){}^2}{ (m-2\tilde{n}+4 r)!}.
\end{multline}

\begin{figure}[H]
    \centering
    \includegraphics[width=0.5\linewidth]{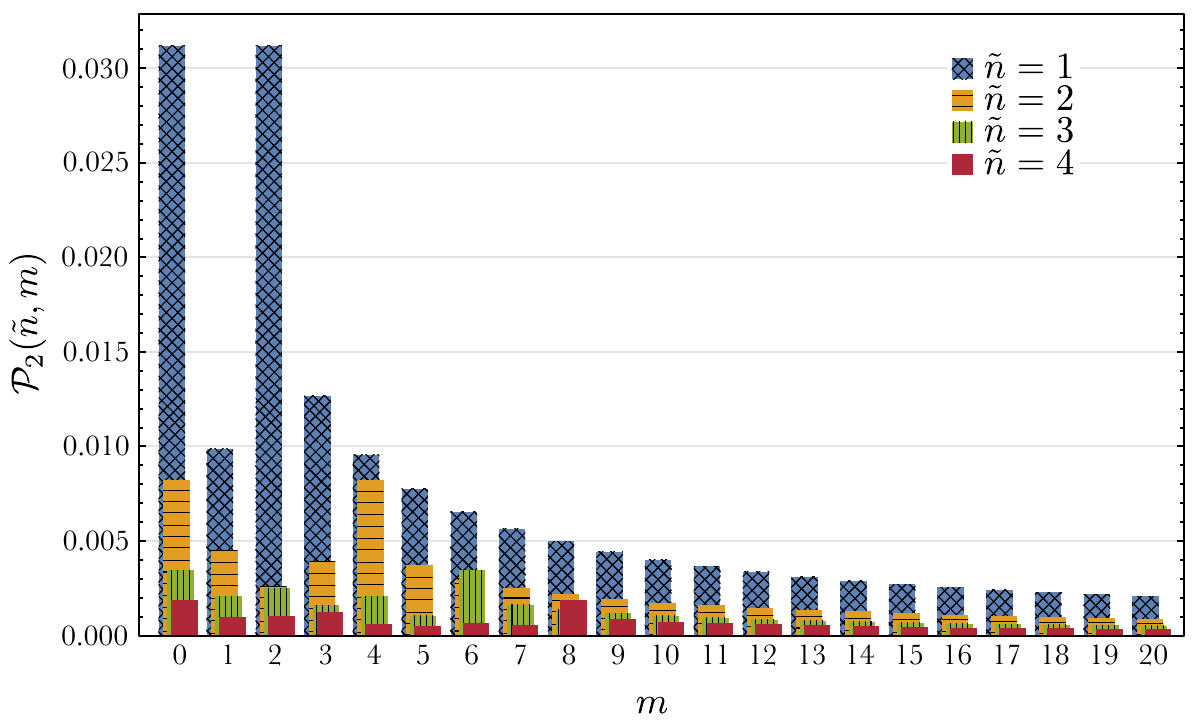}
    \caption{Maximum probability of generating the state (\ref{out_4step_1}) as a function of the number of detected particles $m$. Different colors in the figure denote the cases of measuring different numbers of particles $\tilde{n}$.}
    \label{fig:prob_2step}
\end{figure}

The maximum probability of generating a state at the second iteration of the scheme is given by the expression:
$\mathcal{P}_2(\tilde{n},m)=\max_{a_1,a_2,\tilde{r},\tilde{z}} \left[P_1(\tilde{n},a_1)P_1(\tilde{n},a_2)P_2(\tilde{n},m;\tilde{r},\tilde{z})\right] $.
Fig.  \ref{fig:prob_2step} shows the dependence of the maximum probability of generating the state (\ref{out_4step_1}) on the number of detected particles $m$ for different $\tilde{n}$.

The figure shows that the probability decreases with increasing detected number $m$ and with increasing $\tilde{n}$. Furthermore, the maximum probability is achieved in two cases: $m=0$ and $m=2\tilde{n}$. In these cases, the maximum probabilities coincide and can be calculated analytically. This allows us to write an estimate for the maximum achievable probability of state generation $|\Phi _{\tilde{n},m}^2\rangle$: 
\begin{align} \label{prob_s_21}
    \mathcal{P}_2(\tilde{n},m) \leqslant \left(\frac{\tilde{n}^{\tilde{n}}}{(\tilde{n}+1)^{\tilde{n}+1}}\right)^2 \frac{1}{4^{\tilde{n}}} \binom{2\tilde{n}}{\tilde{n}},
\end{align}
which depends only on the input state parameter $\tilde{n}$.

\begin{figure}[H]
    \centering
    \includegraphics[width=0.5\linewidth]{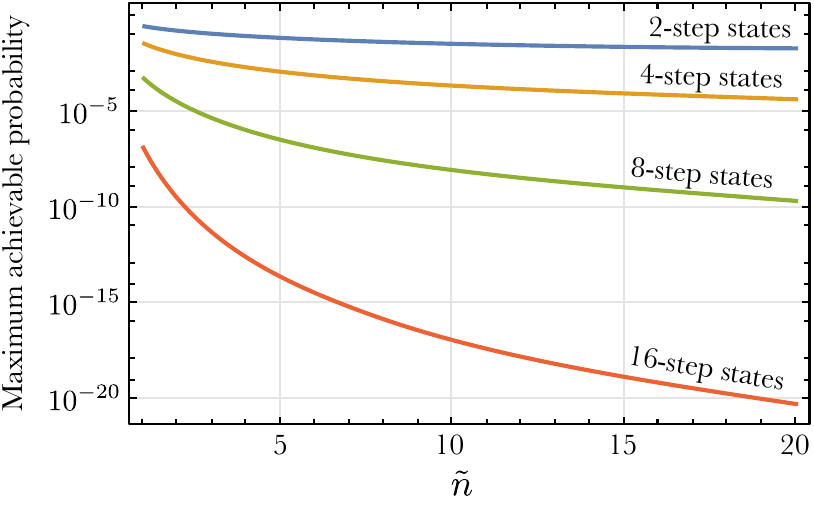}
    \caption{Maximum generation probability for $2^l$-step states as a function of $\tilde{n}$. Different colors indicate states with different step numbers.}
    \label{fig:placeholder1}
\end{figure}

Next, using Eq. (\ref{prob_s_21}), we can find an upper bound on the probability of generating $2^l$-step states:
\begin{align} \label{prob_s_2}
    \mathcal{P}_l(\tilde{n},m_1,\dots,m_{l-1}) \leqslant \left(\frac{\tilde{n}^{\tilde{n}}}{(\tilde{n}+1)^{\tilde{n}+1}}\right)^{2^{l-1}} \left( \frac{1}{4^{\tilde{n}}} \binom{2\tilde{n}}{\tilde{n}}\right)^{2^{l-1}-1}.
\end{align}
For clarity, Fig.  \ref{fig:placeholder1} shows the maximum probability of generating $2^l$-step states as a function of $\tilde{n}$ for different values of $l$.
\noindent Figure shows that the maximum achievable probability decreases rapidly with the number of iterations of the generation scheme. For example, when $\tilde{n}=20$, $16$-step states ($l=4$)  are generated with a probability not exceeding $10^{-20}$, whereas $8$-step states ($l=3$) are generated with a probability not exceeding  $10^{-10}$ at the same $\tilde{n}=20$. The resource states for generating $|\Phi _{\tilde{n},m}^2\rangle$ are squeezed vacuum states, which are produced at a rate of a few GHz. With these states, $8$-step states can be generated within tens of seconds, whereas generating $16$-step states would require prohibitively long times. Thus, at present, realistic experimental generation schemes are limited to producing $8$-step states.

Thus, states with a larger step are potentially better suited for correcting particle-loss errors, but they have a lower generation probability. In realistic schemes, therefore,  we need to find a compromise between the characteristics of the states and the probability of their generation.

\section{Codewords based on 4-step states}

We begin by analyzing the applicability of 4-step states (\ref{out_4step_1}) generated in the considered scheme for correcting particle-loss errors. As CWs, we consider finite superpositions of the same Fock states with different superposition coefficients:
\begin{align}
    &\left|\overline{0}\right\rangle=|\Phi^2_{\tilde{n},m}(\tilde{r}_1,\tilde{z}_2)\rangle=a_0|q\rangle+a_4 |q+4\rangle+a_8 |q+8\rangle, \label{1_CW}\\
    &\left|\overline{1}\right\rangle=|\Phi^2_{\tilde{n},m}(\tilde{r}_2,\tilde{z}_2)\rangle=b_0|q\rangle+b_4 |q+4\rangle+b_8 |q+8\rangle, \label{2_CW}
\end{align}
where $q$ is non-negative integer, and the superposition coefficients are determined by expressions (\ref{coef_4k}). The choice of such CWs is motivated by  several factors. First, these CWs have the same parity (determined by the parity of $q$), which is useful for correcting particle-loss errors. Second, the generated states have two parameters that define the superposition of three Fock states. Adding states to the superposition makes the superposition coefficients mutually dependent.For such states, it is not possible to independently vary all the superposition coefficients. Furthermore, generating a state with a large number of superposition terms  requires measuring a larger number of particles, which, as noted earlier, occurs with a lower probability.


The CWs under consideration, (\ref{1_CW}) and (\ref{2_CW}), depend on the state parameters 
$\tilde r_1$, $ \tilde z_1$ and $\tilde{r}_2$, $\tilde{z}_2$,and on the number of detected particles $(\tilde{n},m)$.  The number $q$ , at which the superposition begins, depends on the numbers $(\tilde{n},m)$, as shown in Eq. (\ref{out_4step_1}).



We choose the state parameters $\lbrace \tilde r_1, \tilde z_1, \tilde{r}_2,\tilde{z}_2 \rbrace $ so that the orthogonality condition for the CWs is satisfied:
\begin{align} \label{cond_1}
    \left\langle \overline{0} |\overline{1}\right\rangle=a_0b_0+a_4b_4+a_8b_8=0,
\end{align}
as well as the condition for the same mean particle numbers in CWs:
\begin{align} \label{cond_2}
    &\left\langle \overline{0} |\hat{n}|\overline{0}\right\rangle=qa_0^2+(q+4) a_4^2+(q+8)a_8^2=\langle n\rangle,\\
    &\left\langle \overline{1} |\hat{n}|\overline{1}\right\rangle=qb_0^2 +(q+4) b_4^2+(q+8)b_8^2=\langle n\rangle.
\end{align} 
The conditions imposed on the CWs are dictated by the KL condition (\ref{KL_ideal}).

\begin{figure}[H]
    \centering
    \includegraphics[width=0.6\linewidth]{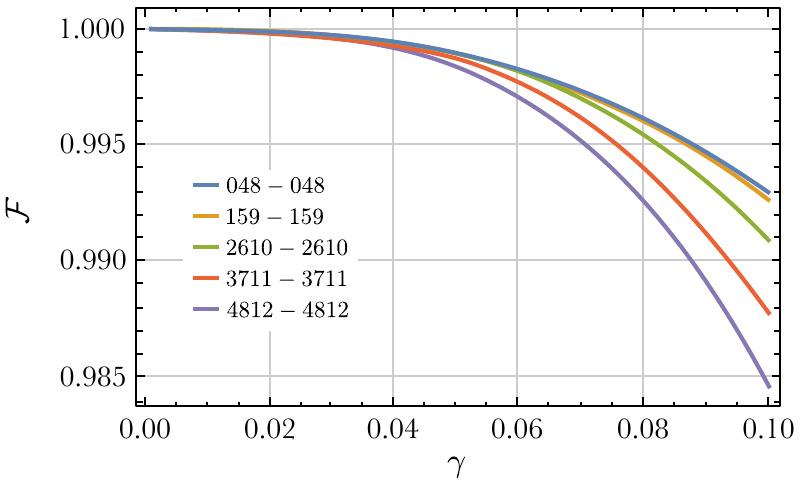}
    \caption{Maximum value of the channel fidelity for CWs (\ref{1_CW}) and (\ref{2_CW}) as a function of the damping parameter $\gamma$. Different colors from top to bottom denote different encodings (different numbers $q$).}
    \label{fig:0-4-8_loss}
\end{figure}

We now assess the efficiency of using the considered CWs for correcting particle-loss errors. As noted earlier, to evaluate the efficiency of error correction with CWs, we use the channel fidelity (\ref{fid_chan}) with Kraus operators for the loss channel (\ref{Krauss_loos}) and the damping parameter $\gamma$. 
The fidelity for CWs (\ref{1_CW}) and (\ref{2_CW}) depends on six parameters $\mathcal{F}=\mathcal{F}(\tilde{r}_1,\tilde{r}_2,\tilde{z}_1,\tilde{z}_2,\tilde{n},m)$. We optimize these parameters to maximize the channel fidelity, subject to the conditions (\ref{cond_1}) and (\ref{cond_2}).
Fig\ref{fig:0-4-8_loss} shows the maximum channel fidelity for the CWs (\ref{1_CW}) and (\ref{2_CW}) as a function of the damping parameter $\gamma$ for different encodings that differ in the lowest Fock state number $q$ in the superposition.

\noindent The figure shows that at low damping parameter $\gamma$, all encodings yield the same fidelity. As $\gamma$ increases, the fidelity decreases at different rates. The channel fidelity decreases faster at higher values of $q$. Such states are less suitable for error correction. This is because as $q$ increases, the mean particle numbers $\langle n\rangle$ in the states also increase. As shown in Ref. \cite{2-step}, the probability of particle-loss for states transmitted through a realistic channel (\ref{Krauss_loos}) depends on the damping parameter $\gamma$ and the mean particle number in the transmitted states. The larger the mean particle number in the transmitted state, the higher the probability of losing several particles. Thus, the best state among those considered is the one whose superposition includes the vacuum state ($q=0$). In what follows, for brevity, we refer to this encoding as \textit{048-048 encoding}.

We now modify the proposed code based on 4-step states. As codewords, we consider arbitrary superpositions of three Fock states with a step of four:
\begin{align}
    &\left|\overline{0}\right\rangle=|\Phi^{1}_{\tilde{n}_1,m_2}(\tilde{r}_1,\tilde z_2)\rangle=a_0|q_1\rangle+a_4 |q_1+4\rangle+a_8 |q_1+8\rangle, \label{1_CW_2}\\
    &\left|\overline{1}\right\rangle=|\Phi^{1}_{\tilde{n}_2,m_2}(\tilde{r}_1,\tilde{z}_2)\rangle=b_0|q_2\rangle+b_4 |q_2+4\rangle+b_8 |q_2+8\rangle, \label{2_CW_2}
\end{align}
where $q_1$ and $q_2$ are non-negative integers, which may differ in general, and the superposition coefficients are given by Eq. (\ref{coef_4k}). As before, we impose two additional conditions on the CWs: orthogonality and equality of the mean particle numbers.

After analyzing the channel  fidelities for various $q_1$ and $q_2$, we find that, among all cases considered, the best encoding has the form:
\begin{align}
    &\left|\overline{0}\right\rangle=a_0|0\rangle+a_4 |4\rangle+a_8 |8\rangle,\\
    &\left|\overline{1}\right\rangle=b_0|2\rangle+b_4 |6\rangle.
\end{align}

We refer to this encoding as \textit{048-26 encoding}. The advantage of this encoding over the others is due to several factors. First, these states are already orthogonal, which means that the orthogonality condition is satisfied automatically, and one of the parameters $\lbrace \tilde{r}_1,\tilde{r}_2,\tilde{z}_1,\tilde{z}_2\rbrace$ becomes free and can participate only in maximizing the fidelity. Second, in this encoding, the smallest distance between CWs (the distance between any two Fock states in the codeword decomposition) is two. This is the maximum distance in encodings based on 4-step states. Such a distance means that the two encoded basis states $|\overline{0}\rangle$ and $|\overline{1}\rangle$ can be confused only if two particle-loss errors occur. Third, this encoding limits the mean particle numbers in the states. Among all possible encodings based on 4-step states, this encoding has the smallest mean particle number, $\langle n\rangle \in [2,6]$.

\begin{figure}[H]
    \centering
    \includegraphics[width=0.6\linewidth]{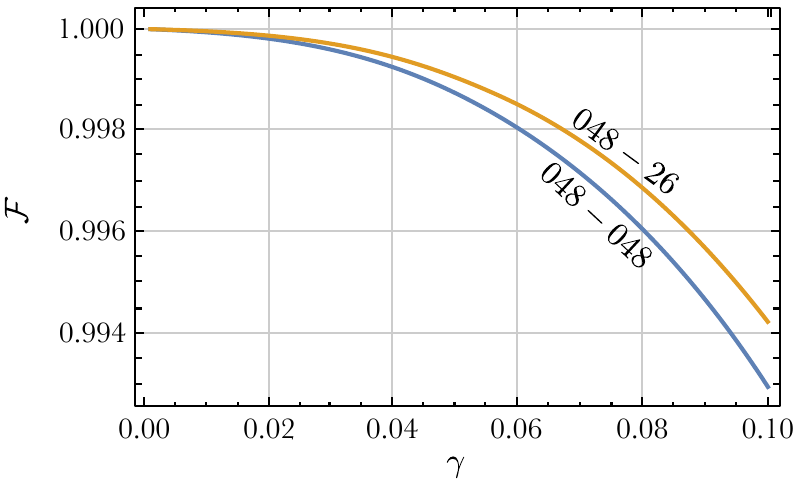}
    \caption{Channel fidelities for the 048-048 and 048-26 encodings as a function of the damping parameter $\gamma$.}
    \label{fig:placeholder2}
\end{figure}

We construct a channel fidelity graph for this encoding in comparison with the best encoding we found earlier to demonstrate the superiority. The graph is presented in Fig. \ref{fig:placeholder2}
\noindent The figure shows that 048-26 encoding is superior to 048-048 encoding. The greatest gain is achieved in the large-$\gamma$ region, since in this region the probability of losing two or more particles is no longer negligible. Thus, we have found the best encoding that uses 4-step states generated in the schemes presented above.

\section{Generalized codewords based on $2^{l}$-step states }
We now generalize the 048-26 encoding found above to the case of states with a larger step. The codewords are based on a superposition of the following form:
\begin{align} 
    &\left|\overline{0}\right\rangle=a_0|0\rangle+a_{2N} |2N\rangle+a_{4N} |4N\rangle, \label{ches_gen_1} \\
    &\left|\overline{1}\right\rangle=b_{N}|N\rangle+b_{3N} |3N\rangle, \label{ches_gen_2}
\end{align}

\begin{figure}[H]
    \centering
    \includegraphics[width=0.6\linewidth]{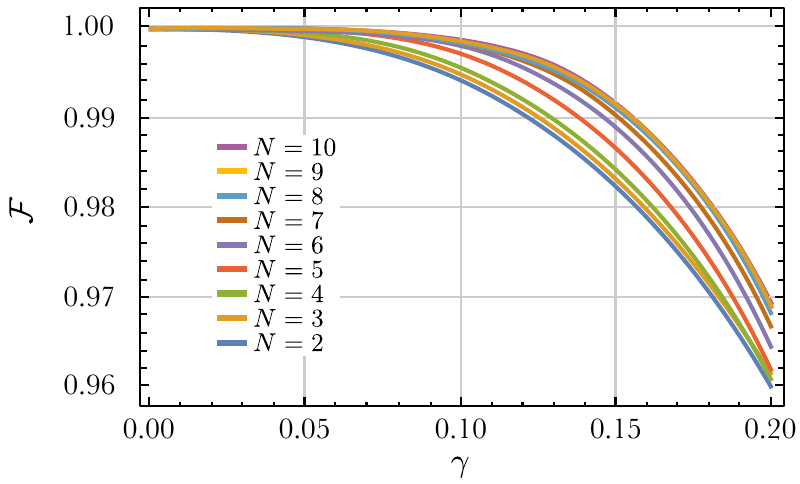}
    \caption{Channel fidelities as a function of the damping parameter $\gamma$ for encodings with different numbers $N$. In the figure, different colors from top to bottom denote different encodings (with different numbers $N$). As $N$ increases, the fidelity also increases.}
    \label{fig:placeholder3}
\end{figure}

We select the superposition coefficients of the states (\ref{ches_gen_1}) and (\ref{ches_gen_2}). This yields CWs that are best suited to protect information in a channel with a damping parameter $\gamma$. Here, for encodings with $N=2,4,8,16,\dots$ we optimize the parameters $\tilde{r}_1$, $\tilde{r}_2$, $\tilde{z}_1$, and $\tilde{z}_2$, rather than the superposition coefficients themselves. These parameters determine the superposition coefficients for the two CWs. We plot the maximum channel fidelity for this encoding as a function of the damping parameter $\gamma$ for different values of $N$. This graph is shown in Fig.  \ref{fig:placeholder3}.
\noindent The figure shows that the channel fidelity increases as $N$ increases. This increase is due to the larger distance between CWs. However, the increase in fidelity is limited. Starting from $N=8$, the channel fidelities become virtually indistinguishable. This is because as $N$ increases, the mean particle numbers in CWs increase, which means the probability of losing several particles in such states also increases. Thus, at large $N$, the benefit of increasing the distance between CWs is offset by the particle-loss rate.

\begin{figure}[H]
    \centering
    \includegraphics[width=0.55\linewidth]{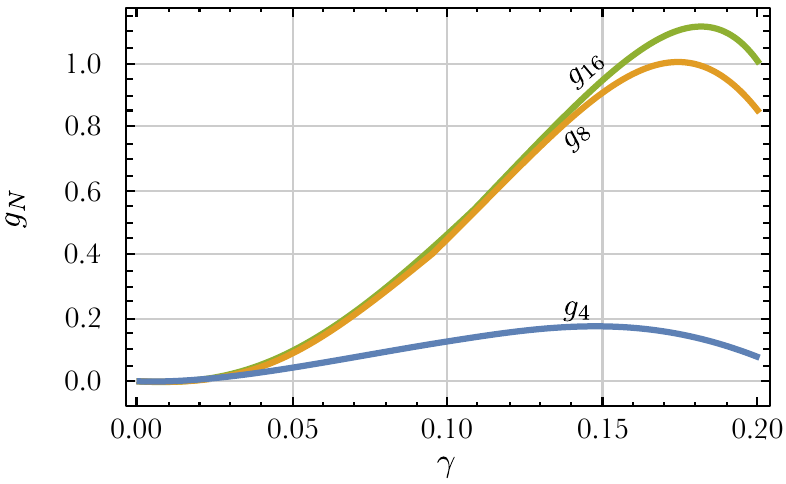}
    \caption{Relative gain in channel fidelity $g_N$ for encodings with $N=4,8,16$ compared with the $N=2$ encoding, as a function of the damping parameter $\gamma$.}
    \label{fig:placeholder4}
\end{figure}

We now consider the channel fidelity for the encodings based on the states generated in our scheme. These are the encodings (\ref{ches_gen_1}) and (\ref{ches_gen_2}) with $N=2,4,8,16$. Fig.  \ref{fig:placeholder4} shows the relative gain in channel fidelity for the encodings $N=4,8,16$, compared to the encoding $N=2$:
$g_N=(\mathcal{F}_{N}/\mathcal{F}_2-1)  \times 100\%$. 
\noindent The figure shows that as $N$ increases, the relative gain increases. However, at $N=4$ ($8$-step states), the gain is insignificant and amounts to less than $0.2 \%$. The maximum gain is achieved with $16$- and $32$-step states ($N=8,16$). However, this gain is about $1\%$. Moreover as shown above, such states are generated with low probability. Thus, we conclude that the best encoding for correcting particle-loss errors is the 048-26 encoding.

\section{Comparison of the proposed encoding with other encodings }
\subsection{Binomial states }
We now compare the performance of our encoding with encodings proposed in other works. We start with the binomial-state encoding \cite{Schlegel2022,Korolev2024_eror_cor, Binomial_state}. This choice is motivated by several factors. First, this encoding can be used to correct particle-loss errors. Second, in this encoding, the CWs are also finite superpositions of Fock states of the form:
\begin{align}
 &|\overline{0}\rangle_{bin}=\frac{1}{\sqrt{2^L}} \sum _{p=0}^{\left\lfloor \frac{L+1}{2}\right\rfloor} \sqrt{\binom{L+1}{2 p}} |2 p (L+1) \rangle,\\
 & |\overline{1}\rangle_{bin}=\frac{1}{\sqrt{2^L}} \sum _{p=0}^{\left\lfloor \frac{L}{2}\right\rfloor} \sqrt{\binom{L+1}{2 p+1}} | (2p+1)(L+1)\rangle.
\end{align}
This encoding depends on the parameter $L$, which determines the maximum number of particle losses that can be corrected by this encoding. For example, for $L=2$, the code can correct the loss of one or two particles.

\begin{figure}[H]
    \centering
    \includegraphics[width=0.5\linewidth]{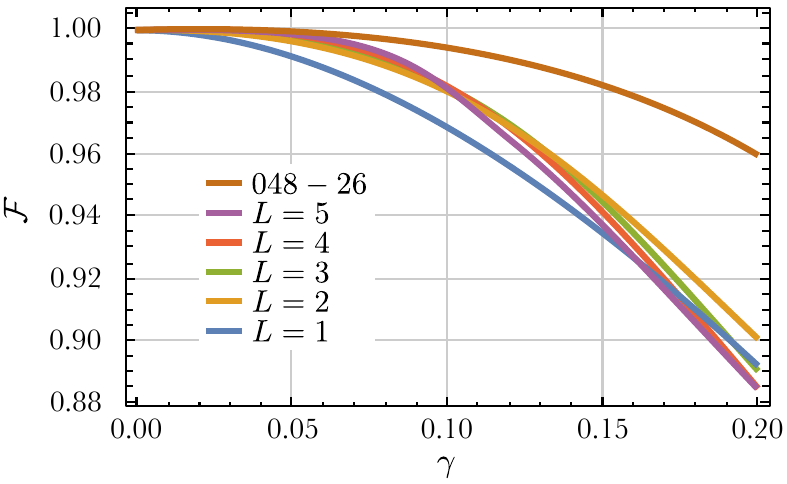}
    \caption{Channel fidelities of the binomial encodings and the 048-26 encoding as functions of the damping parameter $\gamma$. In the figure, different colors from top to bottom denote different encodings: binomial ones with different $L$ and the 048-26 encoding.}
    \label{fig:placeholder5}
\end{figure}

As before, for the binomial states (with various parameters $L$), we calculate the fidelity of the realistic quantum channel with a damping parameter$\gamma$. Fig. \ref{fig:placeholder5} shows these fidelities and the fidelity of our $048-26$ encoding.
\noindent The figure shows that the fidelity of the binomial encoding also saturates as the parameter $L$ increases. The same behavior is observed for our encoding.  Furthermore, the figure indicates that the binomial encoding (even for states with step $2(L+1)=14$) has lower fidelity than 048-26 encoding. The largest difference occurs in the range of large values of the damping parameter $\gamma$. This is because binomial states were not designed for realistic loss channels. They were designed for idealized channels, in which particle-loss is modeled by the annihilation operator. The states proposed here, however, are optimized directly for realistic channels with a given damping parameter $\gamma$.

\subsection{Four-component Schrödinger cat states }
We also compare 048-26 encoding with the encoding based on four-component Schrödinger cat states \cite{Cochrane1999,Bergmann2016,Mirrahimi_2014}. As codewords, we  consider two cat states of the following form:
\begin{align}
&|\overline{0}\rangle_{4cat}\propto|\alpha\rangle+|i\alpha\rangle+|-i\alpha\rangle+|-\alpha\rangle\propto \sum_{j=0}^\infty \frac{\alpha^{4 j}}{\sqrt{(4j)!}}|4j\rangle,\\
 & |\overline{1}\rangle_{4cat}\propto|\beta\rangle-|i\beta\rangle+|-i\beta\rangle-|-\beta\rangle \propto \sum_{j=0}^\infty \frac{\beta^{2+4 j}}{\sqrt{(2+4j)!}}|2+4j\rangle,  
\end{align}
where $|\alpha\rangle$ and $|\beta\rangle$ are coherent states. The comparison of the $048-26$ encoding with this cat encoding is motivated by the similarities between the CWs. First, the four-component Schrodinger cat states have a step 4. Second, as in the  $048-26$ encoding, one of the CWs begins with the vacuum state, and the other with the second Fock state.

\begin{figure}[H]
    \centering
    \includegraphics[width=0.5\linewidth]{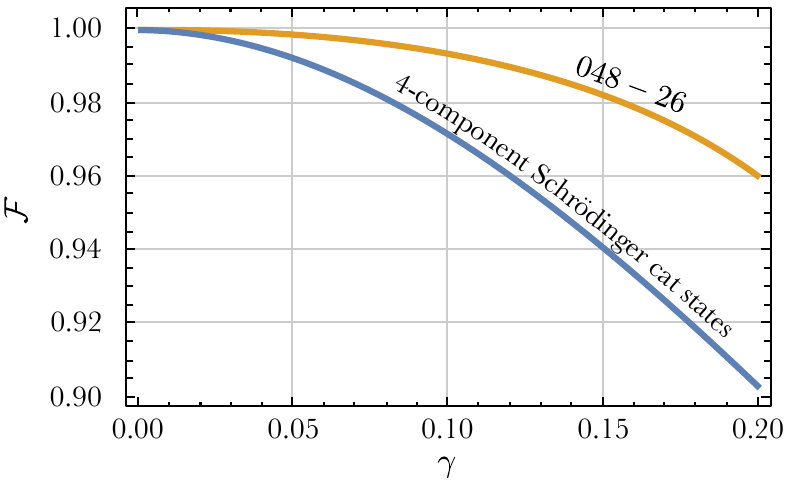}
    \caption{Channel fidelities for the 048-26 encoding and the encoding based on four‑component Schrödinger cat states as functions of the damping parameter $\gamma$.}
    \label{fig:placeholder6}
\end{figure}

In this encoding, we have two parameters $\alpha$ and $\beta$. We choose them  both to equalize the mean particle numbers of the two codewords ${}_{4cat}\langle \overline{0}|\hat{n}|\overline{0}\rangle_{4cat}={}_{4cat}\langle \overline{1}|\hat{n}|\overline{1}\rangle_{4cat}$ and to maximize the fidelity of the quantum channel. Fig.  \ref{fig:placeholder6} shows the maximum fidelity for an encoding based on four‑component Schrödinger cat states, as well as the fidelity of our $048-26$ encoding as a function of the damping parameter $\gamma$.
\noindent The figure shows that the fidelity of the 048-26 encoding exceeds that of the four-component Schrödinger cat encoding across the entire range of $\gamma$. This is because in the 048-26 encoding, each codeword is a superposition of only three Fock states, whereas the cat state contains an infinite number of superposition terms. Moreover, the 048-26 encoding is defined by two parameters, while the cat state depends on only one. These differences significantly complicate the optimization of the cat state to maximize the fidelity of the quantum channel.

\subsection{2-step states}
\begin{figure}[H]
    \centering
    \includegraphics[width=0.5\linewidth]{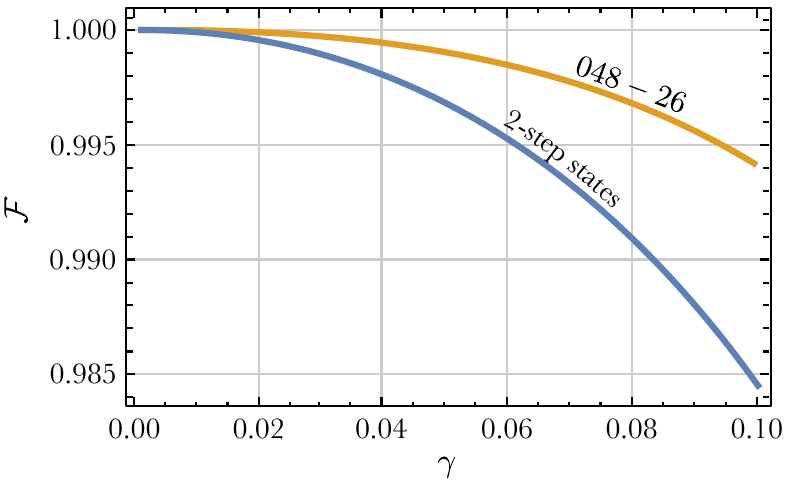}
    \caption{Channel fidelities for the 048-26 encoding and encodings with 2-step states as functions of the damping parameter $\gamma$.}
    \label{fig:placeholder7}
\end{figure}

Finally, we compare the $048-26$ encoding with the encoding based on $|\Phi_n^1(r,z)\rangle$ states . As already mentioned, these states are generated in the scheme presented in Fig. ~\ref{fig:scheme} when two squeezed vacuum states are fed to its input. Ref. \cite{2-step} provides a detailed study of the applicability of such states to particle-loss error correction codes. The states that provided the highest channel fidelity in that study were:
\begin{align}
    &|\overline{0}\rangle _{step2}=|\Psi_n^{out}(r_1,z_1)\rangle, \label{cod_2step_1}\\
    &|\overline{1}\rangle _{step2}=|\Psi_n^{out}(r_2,z_2)\rangle, \label{cod_2step_2}
\end{align}
generated with the same detected particle number  $n$. 
The state parameters $r_1$, $r_2$, $z_1$, and $z_2$ are chosen to satisfy two conditions. The first is the orthogonality condition for the codewords $\langle \overline{0}|\overline{1}\rangle=0$. The second is the equality of the mean particle numbers $\left\langle \overline{0} |\hat{n}|\overline{0}\right\rangle=\left\langle \overline{1} |\hat{n}|\overline{1}\right\rangle$. After these conditions are satisfied, the remaining free parameters are selected to maximize the fidelity of a specific loss channel characterized by a damping parameter $\gamma$. Thus, this encoding is also tailored to a specific damping parameter $\gamma$. Ref.\cite{2-step} demonstrates that the encoding (\ref{cod_2step_1}) and (\ref{cod_2step_2}) with $n=6$ are the best for correcting particle-loss errors. Fig. \ref{fig:placeholder7} shows the fidelity for this encoding, as well as for encoding $048-26$, as a function of the damping parameter $\gamma$.
\noindent  The figure shows that the $048-26$ encoding outperforms the encoding considered above. This is due to the larger step in the codewords. In addition, the larger number of free parameters provides greater flexibility in controlling the codewords.
\section{Conclusion}
The possibility of generating non-Gaussian quantum states of light suitable for protecting quantum information from particle-loss errors is investigated. A scheme based on mixing two input states at a beam splitter, followed by measuring the particle number in one of the modes using a particle-number-resolving detector, is considered. It is shown that squeezed vacuum states used as resource states allow the generation of finite superpositions of Fock states with step two. Successive application of a similar procedure to the resulting states doubles the superposition step at each iteration. This produces states with step equal to a power of two.

The probability of generating states at various iterations of the considered scheme is analyzed. It is shown that increasing the superposition step leads to a rapid decrease in the probability of successful generation. This result establishes a trade-off between the properties of the generated states and the probability of obtaining them. This trade-off must be taken into account when searching for the states best suited for correcting particle-loss errors.

Based on superpositions of three Fock states with step 4, various codeword variants for correcting particle-loss errors are investigated. After analyzing the fidelities for various encodings, the 048-26 encoding is found, which provides the highest fidelity among the considered encodings.

This encoding is generalized to the case of a larger superposition step. It is shown that increasing the step leads to higher fidelity, but this increase is limited. States with large steps have an extremely low probability of generation, while the gain in fidelity provided by small steps is insignificant. Therefore, the best experimentally feasible encoding is the 048-26 encoding.

Finally, the 048-26 encoding is compared with three alternative encodings: binomial codes, four-component Schrödinger cat states, and an encoding based on finite superpositions of squeezed Fock states with step two. It is shown that the proposed encoding provides higher channel fidelity than the alternative encodings studied. Thus, the analysis shows that the proposed 048-26 encoding is promising for experimentally feasible protocols protecting quantum information from particle-loss errors.

This research was supported by the Theoretical Physics and Mathematics Advancement Foundation "BASIS" (Grant No. 24-1-3-14-1).

\bibliography{bibliography}
\end{document}